\documentclass[10pt,twocolumn,twoside]{ieeeconf}
\IEEEoverridecommandlockouts    
\newtheorem{remark}{Remark}

\newtheorem{lemma}{Lemma}
\newtheorem{prop}{Proposition}

\newtheorem{theorem}{Theorem}
\newtheorem{assumption}{Assumption}

\usepackage{cite}
\usepackage{amsmath}
\usepackage{amstext}
\usepackage{amssymb}
\usepackage{tikz}
\usepackage[mathscr]{euscript}

\usepackage{amsfonts}
\usepackage{enumerate}
\usepackage[final]{pdfpages}
\usepackage{csquotes}

\let\labelindent\relax
\DeclareOldFontCommand{\rm}{\normalfont\rmfamily}{\mathrm}

\usepackage{graphicx}      
\usepackage{subcaption}
\usepackage{enumitem}
\usepackage{mathdots}

\usepackage{mathtools,xparse}
\DeclareMathOperator{\diag}{diag}
\definecolor{ForestGreen}{RGB}{34,139,34}
\newcommand{\seb}[1]{%
{\leavevmode\color{black}#1}%
}

\DeclareMathOperator{\sign}{sign}
\def\qed{\hfill $\Box$}

\begin{document}
\author{Sebin~Gracy, Ji~Liu, Tamer Ba\c sar, and C\'esar A.~Uribe
\thanks{Sebin Gracy and C\'esar A.~Uribe are with the Department of Electrical and Computer Engineering, Rice University, Houston, TX, USA (\texttt{sebin.gracy@rice.edu}, \texttt{cauribe@rice.edu}). Ji Liu is with the Department of Electrical and Computer Engineering, Stony Brook University, Stony Brook, NY, USA (\texttt{ji.liu@stonybrook.edu}). Tamer Ba\c sar is with the Coordinated Science Laboratory, University of Illinois  Urbana-Champaign (\texttt{basar1@illinois.edu}).
}}
\title{\LARGE \bf A Discrete-time Networked Competitive Bivirus SIS Model}

\maketitle
\begin{abstract}
The paper deals with the analysis of a discrete-time networked competitive bivirus susceptible-infected-susceptible (SIS) model. More specifically, we suppose that virus~1 and virus~2 are circulating in the population and are in competition with each other. 
 We show that the model is strongly monotone, and that, under certain assumptions, it does not admit any periodic orbit. 
We identify a sufficient condition for exponential convergence to the disease-free equilibrium (DFE). Assuming only virus~$1$ (resp.~virus~$2$) is alive, we establish a condition for global asymptotic convergence to the single-virus endemic equilibrium of virus~1 (resp.~virus~2) - our proof does not rely on the construction of a Lyapunov function.  Assuming both virus~$1$ and virus~$2$ are alive, we establish a condition which ensures local exponential convergence to the single-virus equilibrium of virus~1 (resp. virus~2). Finally, we provide a sufficient (resp. necessary) condition for the existence of a coexistence equilibrium.
\end{abstract}
\section{Introduction}
Mathematical modeling of the spread of an epidemic is extremely advantageous  \cite{hethcote1991modeling}. Indeed, first of all, mathematical models leave no scope for ambiguity by clearly stating the assumptions involved, values assigned to various parameters, etc.  Consequently, these provide conceptual results. Secondly, along with computer simulations, such models  are useful tools for making theoretical advancements and testing their efficacy, developing and checking quantitative conjectures,  determining sensitivities to changes in parameter values, and
estimating key parameters from data. 
Thus, such models contribute towards developing a deeper understanding of how infectious diseases are transmitted in communities, districts, cities, and countries. As a consequence, they can inform better
approaches to eradicating (or, at the very least, mitigating) the spread. Thirdly, they
could also be used for comparing, planning, implementing, and evaluating various
detection, prevention,  and control programs. Lastly, they contribute to the design and analysis of epidemiological surveys, suggest crucial data
that should be collected, identify trends, make general forecasts, etc. \cite{hethcote1991modeling}. \\ 
 \par Over the last several decades, modeling and analysis  of spreading processes has attracted the attention of researchers across a wide spectrum ranging from mathematical epidemiology  \cite{hethcote2000mathematics} and physics \cite{van2009virus} to  the social sciences \cite{easley2010networks}.
Various models have been studied in the literature; see \cite{pare2020modeling} for a recent overview.
This paper focuses on susceptible-infected-susceptible (SIS) models. 

While the (networked) SIS model has been studied in detail (see, for instance, \cite{lajmanovich1976deterministic,gracy2020analysis}), it is not suitable for studying scenarios where there are multiple competing, viruses circulating in the population - a scenario that has been witnessed in the context of spread of gonorrhea and tuberculosis. In the competitive spreading regime, two viruses, say virus~$1$ and virus~$2$, simultaneously circulate in the same population - an individual can either be infected with virus~$1$ or with virus~$2$ or with neither, but not with both. 
Competitive bivirus SIS models have been proposed since \cite{carlos2,castillo1989epidemiological} and more recently in, to cite a few, \cite{sahneh2014competitive,liu2019analysis,ye2021convergence,pare2020analysis,pare2021multi}. The bulk of the literature on networked competitive bivirus  SIS models are focused on the continuous-time case; with the notable exception of \cite{pare2020analysis} (whose analysis of endemic behavior is restricted to providing a lower bound on the number of equilibria) not much attention has been given to the discrete-time networked competitive bivirus  SIS model. The present paper aims to address this gap, specifically by addressing what kinds of behavior the aforementioned model exhibits and also by  shedding more light on the endemic behavior of the same. In more detail, for the discrete-time networked competitive bivirus  SIS model, the paper makes the following contributions: \vspace{-1mm}
\begin{enumerate}[label=\roman*)]
\item We show that the model is strongly monotone, and that, under certain assumptions, it does not admit any periodic orbit; see Proposition~\ref{prop:monotone} and Theorem~\ref{thm:generic:convergence}, respectively.
\item We provide a condition which guarantees exponential convergence to the disease-free equilibrium (DFE); see Theorem~\ref{thm:DFE:expo}.
\item Assuming that only virus~$1$ (resp. virus~$2$) is alive, we secure a condition guaranteeing that for any non-zero initial infection levels the dynamics would converge to the single-virus endemic equilibrium of virus~$1$ (resp. virus~$2$); see Theorem~\ref{thm:boundary:global}. The proof of Theorem~\ref{thm:boundary:global}, unlike that of the single-virus case in \cite{liu2020stability}, does not rely on the construction of an appropriate Lyapunov function.
\item Assuming that both virus~$1$ and virus~$2$ are alive, we identify a condition for local exponential convergence to the single-virus endemic equilibrium of virus~$1$ (resp. virus~$2$); see Theorem~\ref{thm:boundary:local}.
\item We provide a sufficient condition for the existence (resp. nonexistence) of a coexistence equilibrium, i.e., an equilibrium where both viruses are present in a population node; see Theorem~\ref{thm:coexistence} (resp. Theorem~\ref{thm:nece:cond}). 
\end{enumerate}

\subsection*{Paper Outline}
The paper is organized as follows. The notations are listed immediately after the present subsection. The model, technical preliminaries, and formal statements of problems that this paper will investigate are presented in Section~\ref{sec:model}. A condition for global exponential convergence to the DFE is provided in Section~\ref{sec:DFE}, while that for global asymptotic (resp. local exponential)  convergence to the single-virus endemic equilibrium of virus~$1$ (resp. virus~$2$) is given in Section~\ref{sec:analysis:single:virus}. Results on existence (resp. nonexistence) of coexistence equilibirum are provided in Section~\ref{sec:analysis:coexistence}. Numerical examples illustrating our results are provided in Section~\ref{sec;sims}, and finally concluding remarks are given in Section~\ref{sec:conclusion}.

\subsection*{Notations and Preliminaries}
We denote the set of real numbers by $\mathbb{R}$, and the set of nonnegative real numbers by $\mathbb{R}_+$.
 For any positive integer $n$, we use $[n]$ to denote the set $\{1,2,...,n\}$. 
We use $\textbf{0}$ and $\textbf{1}$ to denote the vectors whose entries all equal $0$ and $1$, respectively, and use $I$ to denote the identity matrix,  the sizes of the vectors and matrices are specified only if they are not clear from the context.
 For a vector $x$ we denote the square matrix with $x$ along the diagonal by $\diag(x)$. For any two real vectors $a, b \in \mathbb{R}^n$ we write $a \geq b$ if $a_i \geq b_i$ for all $i \in [n]$, $a>b$ if $a \geq b$ and $a \neq b$, and $a \gg b$ if $a_i > b_i$ for all $i \in [n]$. Likewise, for any two real matrices $A, B \in \mathbb{R}^{n \times m}$, we write $A \geq B$ if $A_{ij} \geq B_{ij}$ for all $i \in [n]$, $j \in [m]$, and $A>B$ if $A \geq B$ and $A \neq B$. 
For a square matrix $M$, we use $\sigma(M)$ to denote the spectrum of $M$, $\rho(M)$ to denote the spectral radius of $M$, and $s(M)$ to denote the largest real part among the eigenvalues of $M$, i.e., $s(M) = \max\{\rm{Re}(\lambda) : \lambda \in \sigma(M)\}$. For a set $\mathcal M$ with boundary, we denote the boundary as $\partial \mathcal M$, and the interior as $\text{Int}(\mathcal M):= \mathcal M\setminus \partial \mathcal M$. Given a matrix $A$, $A \prec 0$ (resp. $ A\preccurlyeq 0 $) indicates that $A$ is negative definite (resp. negative semidefinite), whereas $A \succ 0$ (resp. $ A\succcurlyeq 0 $) indicates that $A$ is positive definite (resp. positive semidefinite). 
A real square matrix $A$ is called Metzler if all its off-diagonal entries are nonnegative.
 A matrix $A$ is said to be an M-matrix if all of its off-diagonal entries are nonpositive, and there exists a constant $c>0$ such that, for some nonnegative $B$ and $c \geq \rho(B)$, $A=cI-B$. All eigenvalues of an M-matrix have nonnegative real parts. Furthermore, if an M-matrix has an eigenvalue at the origin, we say it is singular; if each eigenvalue has strictly positive parts, then we say it is nonsingular. If $A(=[a_{ij}]_{n\times n})$ is a nonnegative   matrix, then $\rho(A)$  decreases monotonically with a decrease in $a_{ij}$ for  any
 $i,j \in [n]$. 

\section{Problem Formulation}\label{sec:model}
In this section, we first introduce the discrete-time networked competitive bivirus SIS model, which is followed by assumptions that are either needed for ensuring that the model is well-defined and/or for paving the way for the main theoretical findings of the present paper. Finally, we provide formal statements of problems that the present paper will focus on. 
\subsection{Model}
We consider two competing viruses, say virus~$1$ and virus~$2$, spreading over a network of $n$ population nodes. Each node is a collection of individuals, and has its own healing (resp. infection) rates with respect to virus~$\ell$, $\delta_i^\ell$ (resp. $\beta_i^\ell$), for $\ell=1,2$. All individuals within a node have the same infection (resp. healing) rates; individuals across different nodes possibly have different infection (resp. healing) rates - that is, homogeneity within a population and heterogeneity across the meta-population. The spread of the two viruses can be represented by a 2-layer graph, say $\mathcal G$. The vertex set of $\mathcal G$ is the set of population nodes; for $\ell=1,2$, the edge set $E^\ell$ captures the interconnection between the various nodes in the context of the spread of virus~$\ell$. We denote by $A^\ell =[a_{ij^\ell}]_{n \times n}$ (where $a_{ij}^\ell \geq 0$) the weighted adjacency matrix for layer $\ell$. Note that $(i,j) \in E^\ell$ if, and only if, $a_{ij}^\ell > 0$. 

We use $x_i^\ell(t)$ to denote the fraction of the population in node $i$ that is infected with virus~$\ell$ at time $t$. The evolution of this fraction is represented by the following scalar differential equation: 
\begin{equation} \label{eq:scalar}
   \dot{x}_i^\ell(t) = - \delta_i^\ell x_i^\ell(t) + \big{(} 1 - \textstyle \sum_{r=1}^m x_i^r(t) \big{)} 
  \textstyle \sum_{j=1}^{n} \beta_{ij}^\ell x_j^\ell(t),
  \end{equation}
where $\beta_{ij}^\ell = \beta_i^\ell a_{ij}^\ell$, and $\ell=1,2$. In vector form, equation~\eqref{eq:scalar} can be written as
\begin{equation}\label{eq:bivirus}
\begin{array}{rcl}
   \dot{x}^1(t) &=  \Big{(} \big{(} I - (X^1+X^2) \big{)} B^1 - D^1 \Big{)} x^1(t), \\
      \dot{x}^2(t) &=  \Big{(} \big{(} I - (X^1+X^2) \big{)} B^2 - D^2 \Big{)} x^2(t), 
\end{array} 
\end{equation} 
where $x^1, x^2 \in \mathbb{R}^n$; $D^\ell, B^\ell$ for $\ell=1,2$ are of appropriate dimensions, and $X^\ell=\diag{(x^\ell)}$ for  $\ell=1,2$.

 The goal of this paper is to consider a discretized version of~\eqref{eq:bivirus}; comment on its limiting behavior above the epidemic threshold; and analyze its various equilibria, viz. existence, uniqueness and  stability. With respect to the former aspect, the present paper aims to develop and gather a series of results that could be viewed as the discrete-time counterparts of (possibly a subset of) the findings in \cite{ye2021convergence,anderson2023equilibria}.\\

\noindent  The discrete-time competitive networked bivirus SIS model that the present paper focuses on is inspired from \cite{pare2020analysis}. Specifically, by applying Euler's forward discretization \cite{atkinson1991introduction} to~\eqref{eq:bivirus}, we obtain the following:
\begin{equation}\label{eq:bivirus:dt}
\footnotesize
\begin{array}{rcl}
   x^1(k+1) &= x^1(k)+ h \Big{(} \big{(} I - (X^1+X^2) \big{)} B^1 - D^1 \Big{)} x^1(k), \\
      x^2(k+1) &= x^2(k)+ h\Big{(} \big{(} I - (X^1+X^2) \big{)} B^2 - D^2 \Big{)} x^2(k).
\end{array} 
\end{equation}

\subsection{Assumptions}
We need the following assumptions so as to ensure that our model is well-defined.
\begin{assumption}\label{assum:1}
For all $i \in [n]$, and $\ell \in [2]$, $x_i^\ell(0)$, $(1-x_i^1(0)-x_i^2(0)) \in [0,1]$.
\end{assumption}

\begin{assumption}\label{assum:2}
For all $i \in [n]$, and $\ell \in [2]$, we have $\delta_i^\ell>0$ and $\beta_{ij}^\ell \geq 0$.
\end{assumption}

\begin{assumption}\label{assum:3}
For all $i \in [n]$, and $\ell \in [2]$, $h\delta_i^\ell<1$ and $h\sum_{\ell=1}^2\sum_{j=1}^{n}\beta_{ij}^\ell\leq 1$.
\end{assumption}

\noindent We define the set $\mathcal D$ as follows: \begin{equation}\label{eq:D}\mathcal D: = \{(x^1, x^2)\mid x^\ell \geq \textbf{0}, \ell=1,2, \textstyle\sum_{\ell=1}^2x^\ell\leq \mathbf{1}\}.\end{equation}

\noindent With Assumptions~\ref{assum:1}-\ref{assum:3} in place, we recall the following.
\begin{lemma}\label{lem:pos:inv}\cite[Lemma~1]{pare2020analysis}
Consider system~\eqref{eq:bivirus:dt} under Assumptions~\ref{assum:1}-\ref{assum:3}. For all $i \in [n]$, and $\ell \in [2]$, $x_i^\ell(k)$, $(1-x_i^1(k)-x_i^2(k)) \in [0,1]$ for all $k \geq 0$.
\end{lemma}

Lemma~\ref{lem:pos:inv} implies that the set $\mathcal D$ is positively invariant. That is, supposing an initial state is in $\mathcal D$, then the forward orbits generated by said initial condition will lie in $\mathcal D$. In other words, Lemma~\ref{lem:pos:inv} ensures that the model in system~\ref{eq:bivirus:dt} is well-defined, in the sense that the state values stay in the interval $[0,1]$ for all time instants; otherwise, since the states represent fractions or approximations of probability, the state values will not correspond to physical reality. Throughout this paper, the term "global" will refer to the following: for all initial conditions in the set $\mathcal D$.\\

\noindent We need the following assumptions for aiding the development of the main results of the present paper.
\begin{assumption}\label{assum:4}
    We have $B^\ell \neq 0$, for each $\ell \in [2]$, $h \neq 0$, and $n > 1$.
\end{assumption}
\noindent Assumption~\ref{assum:4} ensures that we are considering group models, and that there is at least one pair of nodes that share an edge in layer $\ell$ for $\ell =1,2$; otherwise, $B^\ell=0$ for at least one $\ell \in [2]$. Consequently, we are assured that, assuming virus~$1$ (resp. virus~$2$) is present in node $i$ (resp. $j$) for some $i(\text{resp. }j) \in [n]$, the spread is  non-trivial.  
\begin{assumption}\label{assum:5}
    The matrix $B^\ell$ is irreducible, for $\ell=1,2$.
\end{assumption}
\noindent Assumption~\ref{assum:5} is equivalent to insisting that each layer of the spread graph be strongly connected\footnote{A graph is said to be strongly connected if for any pair of nodes $(i,j)$, there exists a path from $i$ to $j$.}.

We need a slightly restrictive version of Assumption~\ref{assum:3}, presented below.
\begin{assumption}\label{assum:6}
For all $i \in [n]$, and $\ell \in [2]$, $h\delta_i^\ell + h\sum_{\ell=1}^2\sum_{j=1}^{n}\beta_{ij}^\ell\leq1$. 
\end{assumption}
\noindent It is immediate that Assumption~\ref{assum:6} implies Assumption~\ref{assum:3}; the converse is not necessarily true.

\noindent System~\eqref{eq:bivirus:dt} has three kinds of equilibria, viz. healthy state or disease-free equilibrium (DFE), $(\textbf{0}, \textbf{0})$; the single-virus endemic equilibrium corresponding to virus~$\ell$ (for each $\ell \in [2]$), $(\bar{x}^\ell, \textbf{0})$, where $\textbf{0}\ll \bar{x}^\ell\ll \textbf{1}$ for $\ell=1,2$; and coexistence equilibria, $(\bar{x}^1, \bar{x}^2)$, where $\textbf{0}\ll \bar{x}^1, \bar{x}^2 \ll \textbf{1}$, and, furthermore, $\bar{x}^1+ \bar{x}^2 \ll \textbf{1}$.  The Jacobian associated with system~\eqref{eq:bivirus:dt}, evaluated at an arbitrary point  $(x^1,x^2)$ in the state space, is as given in~\eqref{jacob}.
\begin{figure*}[h!]
\begin{equation}\label{jacob}
J(x^1,x^2) = \\
\scriptsize
\begin{bmatrix}
I-hD^1+h(I-X^1-X^2)B^1-h\diag(B^1x^1) & -h\diag(B^1x^1)    \\
-h\diag(B^2x^2) & I-hD^2+h(I-X^1-X^2)B^2-h\diag(B^2x^2) \end{bmatrix}
\end{equation}
\end{figure*}

\subsection{Technical preliminaries}
\noindent We will  be needing the following technical details in the sequel \cite{hirsch2005monotone,deplano2020nonlinear}.
A continuous map $T : X \rightarrow X$ on the subset $X \subset Y$ is
\begin{enumerate}[label=\roman*)]
\item monotone if, for any $x, y \in X$,  $x \leq y \implies Tx \leq Ty$
\item strongly monotone if $x < y \implies Tx \ll Ty$
\item strongly order-preserving (SOP) if T is monotone, and when $x < y$ there exist respective neighborhoods $U, V$ of $x, y$ and $n_0 \geq 1$ such that $n \geq n_0 \implies T^nU \leq T^nV$.
\item type-K monotone if $\forall x, y \in \mathbb{R}_{\geq 0}^n$ and $x<y$, it follows that for each $i \in [n]$
    \begin{enumerate}
        \item $x_i=y_i$ $\implies$ $f(x_i) \leq f(y_i)$; and
        \item $x_i < y_i$ $\implies$ $f(x_i) < f(y_i)$.
    \end{enumerate}
\end{enumerate}


\noindent Consider the system
\begin{equation}\label{eq:dt:general}
x(k+1) = f(x(k)).
\end{equation}
Throughout, we will assume that $f \in \mathcal C^1$, where $\mathcal C^1$ denotes the class of continuously differentiable functions. Let $J(.)$ denote the Jacobian associated with system~\eqref{eq:dt:general}. We say that system~\eqref{eq:dt:general} is \emph{monotone} if the matrix $J(.)$ has only nonnegative entries irrespective of the argument \cite[page~141]{hirsch2006attractors}; if the matrix $J(.)$ is also irreducible, then we say that system~\eqref{eq:dt:general} is \emph{strongly monotone}.\\

We will also require the notion of sub-homogeneous systems, introduced in \cite{deplano2020nonlinear}. We say that a positive map $f:\mathbb{R}^n \rightarrow \mathbb{R}^n$ is sub-homogeneous if $$\alpha f(x) \leq f(\alpha x), \hspace{2mm} \forall x \in \mathbb{R}_{\geq 0}^n \text{ and } \alpha \in [0,1].$$
\subsection{Problem Statements}
With respect to system~\eqref{eq:bivirus:dt}, we ask the following questions:
\begin{enumerate}[label=\roman*)]
\item \label{q1} What kinds of behavior does this system exhibit?
\item \label{q2} What is a sufficient condition for global exponential convergence to the DFE?
\item \label{q3} What is a sufficient condition for global asymptotic convergence to a single-virus endemic equilibrium?
\item  \label{q4} Can we identify a sufficient condition for the  local exponential stability of the boundary equilibrium?
\item \label{q5} Can we identify a sufficient condition for the existence of a coexistence equilibrium?
\item \label{q6} Can we identify a sufficient condition for the nonexistence of a coexistence equilibrium?
\end{enumerate}

\section{System~\eqref{eq:bivirus:dt} is strongly monotone and does not admit periodic orbits}\label{sec:monotone}
In this section, we  first investigate whether (or not) system~\eqref{eq:bivirus:dt} is monotone, and subsequently leverage the answer to said question to draw overarching conclusions about the typical behavior of the system. We have the following result.
\begin{prop}\label{prop:monotone}
Under Assumptions~\ref{assum:1},\ref{assum:2}, \ref{assum:4}-\ref{assum:6}, system~\eqref{eq:bivirus:dt} is strongly monotone.
\end{prop}
\noindent \textit{Proof:} Define $P:= \begin{bmatrix} I_n && \textbf{0} \\ \textbf{0} && -I_{n}\end{bmatrix}$. Therefore, $PJ(x^1,x^2)P$ is as given in~\eqref{eq:jacob:perm}.
\begin{figure*}
\begin{equation} \scriptsize 
PJ(x^1,x^2)P= \begin{bmatrix}I-hD^1+h(I-X^1-X^2)B^1-h\diag(B^1x^1) & h\diag(B^1x^1)    \\
h\diag(B^2x^2) & I-hD^2+h(I-X^1-X^2)B^2-h\diag(B^2x^2)  \end{bmatrix} \label{eq:jacob:perm}
\end{equation}
\end{figure*}
For any point, $(x^1, x^2)$,  in the state space, $x^1(k)+x^2(k) \ll \textbf{1}$, for all finite $k>0$; the proof for this claim is immediate from \cite[Lemma~2.1]{ye2021convergence}. This implies that, since $h>0$, for any $k>0$, $h(I-X^1-X^2)$ is a positive diagonal matrix. By Assumptions~\ref{assum:2} and~\ref{assum:5}, we have that $B^\ell$ is nonnegative and irreducible, respectively, for $\ell=1,2$; hence, implying that $h(I-X^1-X^2)B^\ell$ is nonnegative irreducible for $\ell=1,2$. Since by Lemma~\ref{lem:pos:inv}, the set $D$ is positively invariant, it follows that $x^\ell(k) \in [0,1]$ for all $k \in \mathbb{Z}_{\geq 0}$. This, coupled with Assumption~\ref{assum:5}, guarantees that the matrix $h\diag(B^\ell x^\ell)$, for $\ell=1,2$, does not have an all-zero row. Furthermore, from Assumption~\ref{assum:6}, it must be that the matrix $I-hD^\ell+h(I-X^1-X^2)B^\ell-h\diag(B^\ell x^\ell)$, for $\ell=1,2$, is nonnegative and irreducible. Therefore, it follows that the matrix $PJ(x^1,x^2)P$ is nonnegative irreducible irrespective of the states $(x^1, x^2)$. Therefore, from \cite[page~70]{sontag2007monotone}, it follows that system~\eqref{eq:bivirus:dt} is strongly monotone.~$\square$

\noindent Proposition~\ref{prop:monotone} can be viewed as not just the discrete-time counterpart of \cite[Lemma~3.3]{ye2021convergence} but also a stronger version of the same, since Proposition~\ref{prop:monotone} establishes that the map which governs the dynamics of system~\eqref{eq:bivirus:dt} is strongly monotone, whereas \cite[Lemma~3.3]{ye2021convergence} only assures that the flow is monotone.

\noindent Proposition~\ref{prop:monotone} should be understood as follows: suppose that $(x_A^1(0), x_A^2(0))$ and $(x_B^1(0), x_B^2(0))$ are two initial conditions in $\textrm{int}(D)$ satisfying i) $x_A^1(0)>x_B^1(0)$ and ii) $x_A^2(0)<x_B^2(0)$. Since system~\eqref{eq:bivirus:dt} monotone, it follows that, for all $k \in \mathbb{Z}_{\geq 0}$, i) $x_A^1(k)\gg x_B^1(k)$ and ii) $x_A^2(k)\ll x_B^2(k)$.\\


\noindent By leveraging the fact that system~\eqref{eq:bivirus:dt} is monotone, we can draw overarching conclusions on the kinds of  behavior that system~\eqref{eq:bivirus:dt} exhibits. Roughly speaking, we are able to say what happens to the trajectories of system~\eqref{eq:bivirus:dt} (corresponding to almost all initial conditions) as time goes to infinity. The formal details are in the next theorem, prior to which we define the following map associated with system~\eqref{eq:bivirus:dt}.
Define \break 
\begin{align}\label{map}
f(x(t)):&= \tiny\begin{bmatrix}
I+h(I-(x^1+X^2)B^1-D^1)&\textbf{0}\\
\textbf{0} & I+h(I-(x^1+X^2)B^2-D^2)
\end{bmatrix} \nonumber \\
&~~~\times x(t)
\end{align} 
\begin{theorem}\label{thm:generic:convergence} Consider system~\eqref{eq:bivirus:dt} under Assumptions~\ref{assum:1},\ref{assum:2}, \ref{assum:4}-\ref{assum:6}. Suppose that there exists a fixed point in $\text{int }\mathcal D$. Then all periodic points of system~\eqref{eq:bivirus:dt} are fixed points. Furthermore, $\lim_{x\to\infty} f^k(x) =\bar{\bar{x}}$ for all $x(0) \in \mathcal D$, where $\bar{\bar{x}}$ is a fixed point of $f(.)$ in $\text{int }\mathcal D$.
\end{theorem}
\noindent \textit{Proof:} From Proposition~\ref{prop:monotone}, it is clear that the map $f(.)$, as defined in~\eqref{map}, is, due to Assumptions~\ref{assum:1},\ref{assum:2}, \ref{assum:4}-\ref{assum:6}, strongly monotone. Consequently, from the definitions of strong monotonicity and type-K monotoncity, it is clear that $f(.)$ is also type-K monotone. It can be immediately verified that system~\eqref{eq:bivirus:dt} is sub-homogeneous. By assumption, the map $f(.)$ has a fixed point in $\text{int }\mathcal D$. Therefore, from \cite[Theorme~13]{deplano2020nonlinear} (also \cite[Theorem~5.7]{hirsch2020positive}), it follows that all periodic points of system~\eqref{eq:bivirus:dt} are fixed points, and $\lim_{x\to\infty} f^k(x) =\bar{\bar{x}}$ for all $x(0) \in \mathcal D$, where $\bar{\bar{x}}$ is a fixed point of $f(.)$ in $\text{int }\mathcal D$.~\qed

\noindent Note that Theorem~\ref{thm:generic:convergence} excludes the possibility of existence of limit cycles. Also, note that given that system~\eqref{eq:bivirus:dt} is monotone, no other complex behavior is allowed; see \cite[page~70]{sontag2007monotone}.

\begin{remark}[Key difference with the continuous-time setting]
Theorem~\ref{thm:generic:convergence} is the discrete-time counterpart of \cite[Theorem~3.6]{ye2021convergence}. 
The crucial difference is that Theorem~\ref{thm:generic:convergence} relies on the assumption that there exists an equilibrium of system~\eqref{eq:bivirus:dt} in the interior of $\mathcal D$, while \cite[Theorem~3.6]{ye2021convergence} does not need such an assumption.
\end{remark}

The result in Theorem~\ref{thm:generic:convergence}, as mentioned previously, relies on the assumption that there exists a fixed point in $\text{int } \mathcal D$. Indeed, system~\eqref{eq:bivirus:dt} admits an equilibrium in $\text{int } \mathcal D$. A parameter-based condition which ensures the admittance of such an eequilibrium has been provided in \cite[Theorem~12]{cui2022discrete}, while another condition will be provided in Theorem~\ref{thm:coexistence} of the present paper.

    

\section{Analysis of the DFE}\label{sec:DFE}
By looking at equation~\eqref{eq:bivirus:dt}, and by invoking the definition of fixed point of a discrete map, it is immediate that the DFE is always an equilibrium point of system~\eqref{eq:bivirus:dt}; this is independent of any conditions that the system parameters may (or may not) fulfil. We recall the following result.

\begin{prop}\label{prop:DFE}\cite[Theorem~1]{pare2020analysis} Consider system~\eqref{eq:bivirus:dt} under Assumptions~\ref{assum:1}-\ref{assum:5}. Suppose that $\rho(I-hD^\ell+hB^\ell)\leq 1$ for $\ell=1,2$. Then, the DFE is asymptotically stable with a domain of attraction $\mathcal D$, where $\mathcal D$ is as defined in~\eqref{eq:D}.
\end{prop}

\noindent It turns out that if the inequalities in Proposition~\ref{prop:DFE} are tightened, then, one obtains exponential convergence to the DFE, as we detail in the following theorem.
\begin{theorem}\label{thm:DFE:expo}
Consider system~\eqref{eq:bivirus:dt} under Assumptions~\ref{assum:1}-\ref{assum:3}. Suppose that $\rho(I-hD^\ell+hB^\ell)< 1$ for $\ell=1,2$. Then, the DFE is exponentially stable with a domain of attraction $\mathcal D$, where $\mathcal D$ is as defined in~\eqref{eq:D}.
\end{theorem}
\noindent The proof is similar to that of \cite[Theorem~1]{sebin:ifac23}; the details are provided in the interest of completeness.\\

\noindent \textit{Proof:} Define $M^\ell:= I-hD^\ell+hB^\ell$ for $\ell=1,2$. Observe that, as  a consequence of Assumptions~\ref{assum:2} and~\ref{assum:5}, the matrix $M^\ell$ is nonnegative irreducible, for $\ell=1,2$. Let $\ell=1$. By assumption, $\rho(M^1)<1$. Therefore, from \cite[Proposition~1]{rantzer2011distributed}, we know that there exists a diagonal matrix $P^1 \succ 0$ such that $(M^1)^{\top} P^1M^1-P^1 \prec 0$.\\
Consider the following Lyapunov  function candidate $V_1(x^1) = (x^1)^{\top} P^{1}x^1$. It is immediate that, since $P^1 > 0$, $V_1(x^1) >0$ for all $x^1 \neq \textbf{0}$. 
Since $P^1 > 0$, it is also symmetric. Therefore, by applying the Rayleigh-Ritz Theorem (RRT) \cite{horn2012matrix}, it follows that $\lambda_{\min}(P^1)I \leq P^1 \leq \lambda_{\max}(P^1)I$, and 
\begin{align}\label{eq:positiveconstants}
\lambda_{\min}(P^1)\left\|x^1\right\|^{2} \leq V_1(x^1) \leq \lambda_{\max}(P^1)\left\|x^1\right\|^{2}.
\end{align}
Observe that since $P^1 >0$, all its eigenvalues are positive; hence, $\lambda_{\min}(P^1)>0$ and $\lambda_{\max}(P^1)>0$. Therefore, it follows from~\eqref{eq:positiveconstants}  that the constants bounding the Lyapunov function candidate are strictly positive. \\
Define $\Delta V_1(x^1(k)): = V_1(x^1(k+1)) - V_1(x^1(k))$. Hence, for all $x^1\neq \textbf{0}$, we have the following: 

\vspace{-3ex}
\scriptsize
\begin{align}
     \Delta V_1(x^1(k))    = &x^1(k+1)^\top P^1 x^1(k+1) - x^1(k)^\top P^1 x^1(k) \nonumber \\
    =& x^1(k)^\top (\hat{M}^1){^\top} P^1 \hat{M}^1 x^1(k) - x^1(k)^\top P^1 x^1(k) \nonumber \\
    =& x^1(k)^\top \big((M^1 -h\textstyle\sum_{\ell=1}^1 \diag(x^\ell)B^1)^\top P^1 (M^1 \nonumber \\&~~~~~~~~-h\textstyle\sum_{\ell=1}^2 \diag(x^\ell)B^1)\big) x^1(k) 
 -x^1(k)^\top P^1 x^1(k) \nonumber \\
    = & (x^1)^\top (M^1){^\top} P^1 M^1 x^1 - (x^1)^\top P^1 x^1 \nonumber \\&
-2h(x^1)^\top\textstyle\sum_{\ell=1}^2 \diag(x^\ell)B^1 P^1 M^1 x^1 \nonumber \\
    &+h^2(x^1)^\top \Big((\textstyle\sum_{\ell=1}^2 \diag(x^\ell)B^1)^\top P^1 \textstyle\sum_{\ell=1}^2 \diag(x^\ell)B^1\Big) x^1.
    \label{eq:lyap:diff}
\end{align}
\normalsize

\noindent Observe that \scriptsize
\begin{align}
   &-2h(x^1)^\top\textstyle\sum_{\ell=1}^2 \diag(x^\ell)B^1P^1M^1x^1 +h^2(x^1)^\top \times \nonumber \\
   &\qquad \qquad \Big((\textstyle\sum_{\ell=1}^2 \diag(x^\ell)B^1)^\top P^1 \textstyle\sum_{\ell=1}^2 \diag(x^\ell)B^1\Big) x^1 \nonumber \\
   &\leq (x^1)^\top \Big(h^2 (B^1)^\top \textstyle\sum_{\ell=1}^2 \diag(x^\ell)P^1 \textstyle\sum_{\ell=1}^2 \diag(x^\ell)B^1 \nonumber \\
   &~~~~~~~~~~~~-2h^2(B^1)^\top P^1 \textstyle\sum_{\ell=1}^2 \diag(x^\ell)B^1\Big)x^1 \label{ineq:ecc1}\\
   &\leq (x^1)^\top h^2 \Big((B^1)^\top \textstyle\sum_{\ell=1}^2 \diag(x^\ell)P^1 \textstyle\sum_{\ell=1}^2 \diag(x^\ell)B^1 \nonumber \\
   &~~~~~~~~~~~~~~~~~~-(B^1)^\top P^1 \textstyle\sum_{\ell=1}^2 \diag(x^\ell)B^1\Big)x^1 \label{ineq:ecc:2}\\
   & =-(x^1)^\top h^2 \Big((B^1)^\top (I-\textstyle\sum_{\ell=1}^2 \diag(x^\ell)P^1 \textstyle\sum_{\ell=1}^2 \diag(x^\ell) B^1\Big)x^1 \nonumber \\
 &\leq 0, \label{ineq:ecc:3}
\end{align} \normalsize 
where inequality~\eqref{ineq:ecc1} comes from noting that i)  due to  Assumption~\ref{assum:2} the matrix $B^1$ is nonnegative, and ii)  due to  Assumption~\ref{assum:3}, the matrix $(I-hD^1)$ is nonnegative. 
Consequently, the term $-2h(x^1)^\top (I-D^1)P^1\textstyle\sum_{\ell=1}^2 \diag(x^\ell)B^1(x^1)$ is nonpositive. Inequality~\eqref{ineq:ecc:2} is a consequence of Assumption~\ref{assum:3}, whereas inequality~\eqref{ineq:ecc:3} follows by extending the argument in \cite[Lemma~6]{axel2020TAC} 
to
the bivirus case. Therefore, from~\eqref{eq:lyap:diff}, it follows that 
\begin{align}
   \Delta V_1(x^1) \leq  (x^1)^\top \big( {M^1}^\top P^1 M^1  -  P^1 \big)x^1. \label{ineq:lyap:bound}
\end{align} 
Since $(M^1)^\top P^1 M^1-P^1$ is negative definite, it follows that $(M^1)^\top P^1 M^1-P^1$ is symmetric. Consequently,  its spectrum is real, and all its eigenvalues are negative. Therefore, by RRT, we have
\begin{align}
     \Delta V_1(x^1)  \leq -\lambda_{\min}(P^1 -(M^1)^\top P^1 M^1)\left\|x^1\right\|^{2}, \label{constant3}
\end{align}
where $\lambda_{\min}(P^1 {-}(M^1)^\top P^1 M^1)>0$. From~\eqref{eq:positiveconstants} and~\eqref{constant3}, we have that  there exists positive constants, $\lambda_{\min}(P^1)$, $\lambda_{\max}(P^1)$, and $\lambda_{\min}(P^1 {-}(M^1)^\top P^1 M^1)$, such that for~\mbox{$x^1 {\neq} \textbf{0}$},  
\begin{align}
\lambda_{\min}(P^1)\left\|x^1\right\|^{2} \leq V_1(x^1) \leq \lambda_{\max}(P^1)\left\|x^1\right\|^{2}, \label{a1}\\
     \Delta V_1(x^1)  \leq -\lambda_{\min}(P^1 -(M^1)^\top P^1 M^1)\left\|x^1\right\|^{2}.\label{a2}
\end{align}
\noindent Therefore, from \cite[Section~5.9 Theorem.~28]{vidyasagar2002nonlinear}, it is clear that $x^1(k) \rightarrow \textbf{0}$ exponentially fast.\\

\noindent Let $\ell=2$. By assumption, $\rho(M^2)<1$. Therefore, from \cite[Proposition~1]{rantzer2011distributed}, we know that there exists a diagonal matrix $P^2 \succ 0$ such that $(M^2)^{\top} P^2M^2-P^2 \prec 0$. 
Consider the following Lyapunov  function candidate $V_2(x^2) = (x^2)^{\top} P^{2}x^2$. By similar analysis as above, subsequent to a suitable adjustment of notation, it can be shown that $x^2(k) \rightarrow \textbf{0}$ exponentially fast.  Therefore, it follows that  the DFE is exponentially stable with a domain of attraction $\mathcal D$, where $\mathcal D$ is as defined in~\eqref{eq:D}.~\qed



\section{Analysis of the single-virus endemic equilibrium}\label{sec:analysis:single:virus}
It is known that the conditions in Proposition~\ref{prop:DFE} guarantees that the DFE is the unique equilibrium of system~\eqref{eq:bivirus:dt}; see \cite[Theorem~2]{pare2020analysis}. If one of these two spectral radii condition are violated, i.e., , if $\rho(I-hD^\ell+hB^\ell)> 1$ for some $\ell \in [2]$, then it turns out that there exists, besides the DFE, the  single-virus endemic equilibrium (also interchangeably referred to as boundary equilibrium) corresponding to virus~$\ell$, namely $(\bar{x}^\ell, \textbf{0})$; see \cite[Proposition~2]{pare2020analysis}. Furthermore, 
$(\bar{x}^\ell, \textbf{0})$ is locally asymptotically stable; see~\cite[Corollary~1]{pare2020analysis}. However, \cite{pare2020analysis} makes no comment on the global asymptotic stability of $(\bar{x}^\ell, \textbf{0})$. In this section, we first strengthen \cite[Corollary~1]{pare2020analysis} by establishing global asymptotic stability of $(\bar{x}^\ell, \textbf{0})$. Second, we allow for $\rho(I-hD^\ell+hB^\ell)> 1$ for each $\ell \in [2]$, and establish local exponential convergence to $(\bar{x}^\ell, \textbf{0})$.
\par It turns out that one can leverage a result on discrete maps from \cite{hirsch2005monotone} 
to guarantee global asymptotic stability of the single-virus endemic equilibrium.
Before presenting the result, we recall the notion of ordered fixed points. Consider am arbitrary map $f(.)$, let $y_1$ and $y_2$ be its fixed points. We say that $y_1$ and $y_2$ are ordered if $y_1 \gg y_2$ or if $y_1 \ll y_2$.
We have the following theorem.
\begin{theorem}\label{thm:boundary:global}
  Consider system~\eqref{eq:bivirus:dt} under Assumptions~\ref{assum:1}-\ref{assum:2}, \ref{assum:4}-\ref{assum:6}. Suppose that $\rho(I-hD^1+hB^1)> 1$ and $\rho(I-hD^2+hB^2)\leq 1$. The boundary equilibrium $(\bar{x}^1, \textbf{0})$ is asymptotically stable,  with a domain of attraction $\mathcal D\setminus \textbf{0}$, where $\mathcal D$ is as defined in~\eqref{eq:D}. 
\end{theorem}
\noindent \textit{Proof:} By assumption, $\rho(I-hD^1+hB^1)> 1$. Therefore, it is immediate that $\rho(J(\textbf{0}, \textbf{0}))>1$, and, hence, the DFE is unstable. Furthermore, from \cite[Proposition~2]{pare2020analysis}, it follows that there exists a boundary equilibrium $(\bar{x}^1, \textbf{0})$, where $\textbf{0} \ll \bar{x}^1 \ll \textbf{1}$. Since, by assumption, $\rho(I-hD^2+hB^2)\leq 1$, it follows that the boundary equilibrium $( \textbf{0}, \bar{x}^2)$, where $\textbf{0} \ll \bar{x}^2 \ll \textbf{1}$ does not exist. Moreover, since $\rho(I-hD^2+hB^2)\leq 1$ (which implies that
virus~$2$ dies out), there can be no equilibria of the form $(\hat{x}^1, \hat{x}^2)$, where $\textbf{0} \ll \hat{x}^1, \hat{x}^2 \ll \textbf{1}$, and $\hat{x}^1+ \hat{x}^2 \ll \textbf{1}$. Therefore, the only possible equilibria of the system are the DFE and $(\bar{x}^1, \textbf{0})$. Hence, there does not exist three fixed points, which further implies that there does not exist three ordered fixed points. Therefore, since we know from Proposition~\ref{prop:monotone} that system~\eqref{eq:bivirus:dt} is monotone, which, by definition, implies that the discrete map is strongly order preserving,  from \cite[Theorem~5.7, statement~iii)]{hirsch2005monotone} (also see \cite{dancer1998some}), it follows that every orbit converges to a fixed point, namely the boundary equilibrium $(\bar{x}^1, \textbf{0})$. Therefore, the boundary equilibrium $(\bar{x}^1, \textbf{0})$ is asymptotically stable, with a domain of attraction $\mathcal D\setminus\textbf{0}$.~$\square$

Theorem~\ref{thm:boundary:global} guarantees global asymptotic stability of the boundary equilibrium $(\bar{x}^1, \textbf{0})$. That is, for all non-zero initial conditions, the dynamics of system~\eqref{eq:bivirus:dt} converge to $(\bar{x}^\ell, \textbf{0})$. 
Note that, for the particular case of single virus spread, \cite[Theorem~1]{liu2020stability} also provides a sufficient condition for GAS of the equilibrium point  $\bar{x}^1$. The proof of \cite[Theorem~1]{liu2020stability} relies on Lyapunov techniques, whereas that of Theorem~\ref{thm:boundary:global} uses results on existence of fixed points in discrete maps, and, as it turns out, is significantly shorter.

Note that Theorem~\ref{thm:boundary:global} allows for, without loss of generality, either $\rho(I-hD^1+hB^1)> 1$ or $\rho(I-hD^2+hB^2)> 1$, but not both. A natural question of interest, then, would be to understand what happens when both  $\rho(I-hD^1+hB^1)> 1$  and $\rho(I-hD^2+hB^2)> 1$. We aim to address the same in the rest of the present paper.

\noindent \begin{theorem}\label{thm:boundary:local}
  Consider system~\eqref{eq:bivirus:dt} under Assumptions~\ref{assum:1}-\ref{assum:2}, \ref{assum:4}-\ref{assum:6}. Suppose that $\rho(I-hD^\ell+hB^\ell)> 1$ for $\ell=1,2$. 
The boundary equilibrium $(\bar{x}^1, \textbf{0})$ is asymptotically stable if $\rho(I-hD^2+(I-\bar{X}^1)B^2) \leq 1$.
  If $\rho(I-hD^2+(I-\bar{X}^1)B^2) > 1$, then the boundary equilibrium $(\bar{x}^1, \textbf{0})$ is unstable.
\end{theorem}
\noindent The proof is inspired from that of \cite[Theorem~3.9]{ye2021convergence}.

\noindent \textit{Proof:} Observe that the Jacobian evaluated at the boundary equilibrium, $(\bar{x}^1, \textbf{0})$,  reads as given in~\eqref{jacob:x1}.
\begin{figure*}
\begin{equation}\label{jacob:x1}
 J(\bar{x}^1, \textbf{0}) = 
\footnotesize  \begin{bmatrix}
I-hD^1+h(I-\bar{X}^1)B^1-h\diag(B^1\bar{x}^1) & -h\diag(B^1\bar{x}^1)    \\
\textbf{0}& I-hD^2+h(I-\bar{X}^1)B^2\end{bmatrix}.
\end{equation}
\end{figure*}
\normalsize
Since $(\bar{x}^1, \textbf{0})$ is an equilibrium point, by definition of equilibrium, it must be that
\begin{equation}\label{key:eq:bounday:stab}
  \Big{(} \big{(} I - \bar{X}^1 \big{)} B^1 - D^1 \Big{)} \bar{x}^1 = \textbf{0}. 
\end{equation}
Note that from \cite[Lemma~6]{axel2020TAC}, it follows that $(I-\Tilde{X}^1)$ is positive diagonal. Therefore, together with Assumptions~\ref{assum:2} and~\ref{assum:6}, it follows that the matrix $(-D^1+(I-\Tilde{X}^1)B^1)$ is irreducible Metzler. Hence, by applying \cite[Lemma~2.3]{varga1999matrix} to~\eqref{key:eq:bounday:stab}, it must be that $\bar{x}^1$ is, up to a scaling, the only eigenvector of $(-D^1+(I-\Bar{X}^1)B^1)$ with all entries being strictly positive. Furthermore, $\bar{x}^1$ is the eigenvector that is associated with, and only with, $s(-D^1+(I-\Bar{X}^1)B^1)$. Therefore,   $s(-D^1+(I-\bar{X}^1)B^1)=0$.\\
Define $Q:=D^1-(I-\Bar{X}^1)B^1$, and note that $Q$ is an M-matrix. Since $s(-Q)=0$ and $\diag(B^1\bar{x}^1)$ is irreducible, it follows that $Q$ is a singular irreducible M-matrix. Observe that $\diag(B^1\bar{x}^1)$ is a nonnegative matrix, and because $B^1$ is irreducible and $\bar{x}^1 \gg \textbf{0}$, it must be that at least one element in $\diag(B^1\bar{x}^1)$ is strictly positive. Therefore, from \cite[Lemma~4.22]{qu2009cooperative}, it follows that $Q+\diag(B^1\bar{x}^1)$ is an irreducible non-singular M-matrix, which from \cite[Section~4.3, page~167]{qu2009cooperative} implies that $-Q-\diag(B^1\bar{x}^1)$ is Hurwitz. Therefore, we have that $s(-D^1+(I-\bar{X}^1)B^1-\diag(B^1\bar{x}^1))<0$. Since $h>0$, $s(-hD^1+h(I-\bar{X}^1)B^1-h\diag(B^1\bar{x}^1))<0$. Hence, from the proof of \cite[Proposition~2]{pare2020analysis}, we have that $\rho(I-hD^1+h(I-\bar{X}^1)B^1-h\diag(B^1\bar{x}^1))<1$. By assumption,  $\rho(I-hD^2+h(I-\bar{X}^1)B^2)<1$. Therefore, since the matrix $J(\bar{x}^1, \textbf{0})$ is block upper triangular, local asymptotic stability is a direct consequence of \cite[page 268, Theorem~42]{vidyasagar2002nonlinear}.


\section{(Non)Existence of a coexistence equilibrium}\label{sec:analysis:coexistence}
The analysis of system~\eqref{eq:bivirus:dt} has as yet focused on the existence and stability of the single-virus endemic equilibria corresponding to virus~$\ell$ for each $\ell \in [2]$. In this section, we aim to provide conditions for the existence (resp. nonexistence) of a (resp. any) coexistence equilibrium.

A sufficient condition for the existence of a  coexistence equilibrium for system~\eqref{eq:bivirus:dt} has been provided in \cite[Theorem~12]{cui2022discrete}. 
Note that \cite[Theorem~12]{cui2022discrete} relies on the assumption that both the boundary equilibria are unstable, i.e., $\rho(I-hD^2+(I-\bar{X}^1)B^2) > 1$ and $\rho(I-hD^1+(I-\bar{X}^2)B^1) > 1$. We establish existence of a coexistence equilibrium for a different stability configuration of the boundary equilibria, namely $\rho(I-hD^2+(I-\bar{X}^1)B^2) < 1$ and $\rho(I-hD^1+(I-\bar{X}^2)B^1) < 1$. 
 We have the following result:
\begin{theorem}\label{thm:coexistence}
     Consider system~\eqref{eq:bivirus:dt} under Assumptions~\ref{assum:1}-\ref{assum:2}, \ref{assum:4}-\ref{assum:6}. Suppose that $\rho(I-hD^\ell+hB^\ell)> 1$ for $\ell=1,2$. Let $(\bar{x}^1, \textbf{0})$ and $(\textbf{0}, \bar{x}^2)$ denote the boundary equilibria corresponding to virus~$1$ and virus~$2$, respectively. 
Suppose that $\rho(I-hD^2+(I-\bar{X}^1)B^2) < 1$ and $\rho(I-hD^1+(I-\bar{X}^2)B^1) < 1$. Suppose further that $\bar{x}^1 < \bar{x}^2$.
Then, there exists an unstable coexistence equilibrium $(\hat{x}^1, \hat{x}^2)$, where $\textbf{0} \ll (\hat{x}^1, \hat{x}^2) \ll \textbf{1}$.
\end{theorem}
\noindent \textit{Proof:} By assumption, $\rho(I-hD^\ell+hB^\ell)> 1$ for $\ell=1,2$. Hence, from \cite[Proposition~2]{pare2020analysis}, it follows that there exists boundary equilibria $(\bar{x}^1, \textbf{0})$ and $(\textbf{0}, \bar{x}^2)$, where $\textbf{0} \ll \bar{x}^1, \bar{x}^2 \ll \textbf{1}$. By assumption, $\rho(I-hD^2+(I-\bar{X}^1)B^2) < 1$ and $\rho(I-hD^1+(I-\bar{X}^2)B^1) < 1$. Consequently, from Theorem~\ref{thm:boundary:local}, it follows that both $(\bar{x}^1, \textbf{0})$ and $(\textbf{0}, \bar{x}^2)$ are stable. Therefore, since 
By assumption $\bar{x}^1 < \bar{x}^2$, and since from Proposition~\ref{prop:monotone} we know that system~\eqref{eq:bivirus:dt} under Assumptions~\ref{assum:1}-\ref{assum:2}, \ref{assum:4}-\ref{assum:6} is strongly monotone, the existence of an unstable coexistence equilibrium, $(\hat{x}^1, \hat{x}^2)$, where $\textbf{0} \ll (\hat{x}^1, \hat{x}^2) \ll \textbf{1}$, follows from \cite[Theorem~4]{hess1991stability}. Furthermore, $(\bar{x}^1, \textbf{0}) <  (\hat{x}^1, \hat{x}^2) < (\textbf{0}, \bar{x}^2)$.

We have the following remark.
\begin{remark}
    Theorem~\ref{thm:coexistence} ensures not just the existence of a coexistence equilibrium but also guarantees that said coexistence equilibrium is unstable. In the context of system~\eqref{eq:bivirus} (which is the continuous-time counterpart of system~\eqref{eq:bivirus:dt}), it is known that the stability configuration of the boundary equilibria as in (the continuous-time version of) Theorem~\ref{thm:coexistence} only ensures that the coexistence equilibrium is either neutrally stable (i.e., for the associated Jacobian, there exists an eigenvalue with real part equal to zero) or unstable; see \cite[Corollary~3.15]{ye2021convergence}. Thanks to \cite{anderson2023equilibria}, where it is shown that for system~\eqref{eq:bivirus} the equilibria are hyperbolic, it is known that generically, i.e., for almost all choices of parameter matrices $D^1, D^2, B^1, B^2$, the coexistence equilibrium is unstable.
\end{remark}

 Note that given a discrete-time bivirus system with dynamics as in~\eqref{eq:bivirus:dt}, it is straightforward to \emph{verify} whether said system fulfills the conditions of Theorem~\ref{thm:coexistence}. The converse problem of \emph{designing} bivirus networks such that the conditions in Theorem~\ref{thm:coexistence} are fulfilled is  more involved; for the continuous-time case, see \cite{ben:pre}.

We identify a sufficient condition for the nonexistence of a coexistence equilibrium.
\begin{theorem}   
\label{thm:nece:cond}
Consider system~\eqref{eq:bivirus:dt} under Assumptions~\ref{assum:1}-\ref{assum:2}, \ref{assum:4}-\ref{assum:6}. Suppose that $\rho(I-hD^\ell+hB^\ell)> 1$ for $\ell=1,2$. Let $(\bar{x}^1, \textbf{0})$ and $(\textbf{0}, \bar{x}^2)$ denote the boundary equilibria corresponding to virus~$1$ and virus~$2$, respectively.  If $\bar{x}^1 \ll \bar{x}^2$, then there does not exist any coexistence equilibrium.
\end{theorem}
\noindent \textit{Proof:} Define $x:=\begin{bmatrix} x^1 \\ x^2\end{bmatrix}$, and observe that the dynamics of system~\eqref{eq:bivirus:dt} can be rewritten as follows:
\begin{align} 
x(k+1)&=\tiny\begin{bmatrix}
I+h(I-(x^1+X^2)B^1-D^1)&\textbf{0}\\
\textbf{0} & I+h(I-(x^1+X^2)B^2-D^2)
\end{bmatrix}  \nonumber  \\ &~~~ \times x(k)
\end{align}
 Hence, it is immediate that $x(k+1) = f(x(k))$, where $f(.)$ is as defined in~\eqref{map}. By Proposition~\ref{prop:monotone}, we know that the mapping $f(.)$ is strongly monotone. Furthermore, since $x(k) \in \mathcal D$  by Lemma~\ref{lem:pos:inv}, where $\mathcal D$ is as defined in~\eqref{eq:D}, for all $k \in \mathbb{Z}_{\geq 0}$, it follows that $f:\mathcal D \rightarrow \mathcal D$. Since the set $\mathcal D$ is closed and bounded, it follows that $f(x)$ has compact closure.  By assumption, $\rho(I-hD^\ell+hB^\ell)> 1$ for $\ell=1,2$. Hence, from \cite[Proposition~2]{pare2020analysis}, it follows that there exist boundary equilibria $(\bar{x}^1, \textbf{0})$ and $(\textbf{0}, \bar{x}^2)$, where $\textbf{0} \ll \bar{x}^1, \bar{x}^2 \ll \textbf{1}$. Therefore, the map $f(.)$ fulfills all the assumptions of the order interval trichotomy theorem; see \cite[Theorem~5.1]{hirsch2005monotone}. Specifically, \cite[Theorem~5.1]{hirsch2005monotone} guarantees that at least one of the following scenarios should happen.
\begin{enumerate}[label=\roman*)]
\item \label{outcome1} there exists a fixed point $\hat{x}$ such that $\bar{x}^1 \ll \hat{x} \ll \bar{x}^2$
\item \label{outcome2} there exists an entire orbit from $\bar{x}^1$ to $\bar{x}^2$ that is increasing
\item \label{outcome3} there exists an entire orbit from $\bar{x}^1$ to $\bar{x}^2$ that is decreasing
\end{enumerate}
Recall that  $\bar{x}^1$ and  $\bar{x}^2$ are fixed points of $f(.)$. By assumption, $\bar{x}^1 \ll \bar{x}^2$. Therefore, from \cite[Proposition~5.3]{hirsch2005monotone}, at most one among scenario~\ref{outcome2} and scenario~\ref{outcome3} happens, which implies that scenario~\ref{outcome1} cannot happen. That is, there  does not exist a fixed point $\hat{x}$ such that $\bar{x}^1 \ll \hat{x} \ll \bar{x}^2$; hence, guaranteeing that there does not exist any coexistence equilibrium.~\qed

For  the continuous-time case, 
if $\bar{x}^2 \gg \bar{x}^1$ then $\bar{x}^2$ is locally exponentially stable, and $\bar{x}^1$ is unstable. while if $B^2>B^1$ then, in addition to the aforementioned stability configuration for the boundary equilibria, it also turns out that there does not exist a coexistence equilibrium; see \cite[Corllary~3.11, statements~1 and 3]{ye2021convergence}. Moreover, it is also known that if $B^2>B^1$ then $\bar{x}^2 \gg \bar{x}^1$, whereas the converse is not necessarily true \cite{ye2021convergence};  this means that for the continuous-time case it is not known if $\bar{x}^2 \gg \bar{x}^1$ implies that there does not exist any coexistence equilibrium. For the discrete-time case, $B^2>B^1$ guarantees the nonexistence of any coexistence equilibrium; see \cite[Theorem~13]{cui2022discrete}, whereas no such result was previously available for the case when $\bar{x}^2 \gg \bar{x}^1$--
Theorem~\ref{thm:nece:cond} closes this gap.

We next identify a condition which ensures that, under the hypothesis of Theorem~\ref{thm:coexistence}, no orbit of system~\eqref{eq:bivirus:dt} converges to the boundary of $\mathcal D$, $\partial\mathcal D$, where $\mathcal D$ is as defined in~\eqref{eq:D}. To this end, we need the following Assumption, which is stronger than Assumption~\ref{assum:5}.
\begin{assumption}\label{assum:7}
        The matrix $B^\ell$ is primitive, for $\ell=1,2$.
\end{assumption}
Observe that every primitive matrix is irreducible; see \cite[Lemma~2.11]{FB-LNS}. Therefore, Assumption~\ref{assum:7} implies Assumption~\ref{assum:5}; the converse is false. 
We have the following result.
\begin{prop}\label{prop:nothing:boundary}
 Consider system~\eqref{eq:bivirus:dt} under Assumptions~\ref{assum:1}, \ref{assum:2}, \ref{assum:4}, \ref{assum:6} and~\ref{assum:7}. Suppose that $\rho(I-hD^\ell+hB^\ell)> 1$ for $\ell=1,2$. There are no orbits remaining in $\partial\mathcal D$.
\end{prop}
\textit{Proof:} Define $x:=\begin{bmatrix} x^1 \\ x^2\end{bmatrix}$, and observe that the dynamics of system~\eqref{eq:bivirus:dt} can be rewritten as follows:
\begin{align} 
x(k+1)&=\tiny\begin{bmatrix}
I+h(I-(x^1+X^2)B^1-D^1)&\textbf{0}\\
\textbf{0} & I+h(I-(X^1+X^2)B^2-D^2)
\end{bmatrix}x(k)
\end{align}
Define $A(x):=\tiny\begin{bmatrix}
I+h(I-(x^1+X^2)B^1-D^1)&\textbf{0}\\
\textbf{0} & I+h(I-(X^1+X^2)B^2-D^2)
\end{bmatrix}$.
It is immediate that, for all $i,j \in [2n]$, $a_{ij}(x)$ is continuous in $x$.
From Assumption~\ref{assum:3} it is clear that $I-hD^\ell$, for $\ell=1,2$, is a positive diagonal matrix. Furthermore, it can be shown that $x^1(k)+x^2(k) \ll \mathbf{1}$ for all $k<\infty$. Also, since by Assumption~\ref{assum:7}, $B^\ell$, for $\ell=1,2$, is primitive, it follows that the matrices $ I+h(I-(X^1+X^2))B^\ell-hD^\ell$ for $\ell=1,2$ are irreducible. Therefore, the matrix $A(x)$ is irreducible for all $x \in \mathcal D$, in particular for all $x \in \partial \mathcal D$. Together with Assumption~\ref{assum:6}, we can conclude that $A(x)$ is nonnegtaive irreducible for all $x \in \mathcal D$. Therefore, by exploiting the fact that $A(x)$ is  irreducible, from \cite[Lemma~1]{liu2019analysis}, it is clear that, for $x\geq\textbf{0}$, $A(x)x \geq \textbf{0}$. Also, since $A(x)$ is nonnegtaive, it is immediate that, for $x>\textbf{0}$, $A(x)x > \textbf{0}$. Lastly, Lemma~\ref{lem:pos:inv} guarantees that system~\eqref{eq:bivirus:dt} is pointwise dissipative.\\
Note that an orbit $\{x(k)\}_{k \in \mathbb{Z}_{\geq 0}}$ could either start in $\text{int }\mathcal D$ or in $\partial\mathcal D$. We will deal with these two cases separately.\\
\textbf{Case 1 ($x(0) \in \text{int }\mathcal D$):} Note that $A(0):=\tiny\begin{bmatrix}
I+hB^1-hD^1&\textbf{0}\\
\textbf{0} & I+hB^2-hD^2
\end{bmatrix}$. Since  $x^1(k)+x^2(k) \ll \mathbf{1}$, it is immediate that $\sign(B^\ell)=\sign((I-X^1-X^2)B^\ell)$ for $\ell=1,2$, which, because of Assumption~\ref{assum:3}, further implies that $\sign(A(0))=\sign(A(x))$ for all $x \in \mathcal D$. Furthermore, by assumption, $\rho(I-hD^\ell+hB^\ell)> 1$ for $\ell=1,2$. Therefore, all the assumptions of \cite[Theorem~4.3]{kon2005nonexistence} are satisfied. Consequently, since, by Assumption~\ref{assum:7}, $A(0)$ is primitive, it follows from \cite[Theorem~4.3]{kon2005nonexistence}  that there are no orbits $\{x(k)\}_{k \in \mathbb{Z}_{\geq 0}}$ starting in 
$\text{int }\mathcal D$ that converges to $\partial\mathcal D$.\\
\textbf{Case 2 ($x(0) \in \partial\mathcal D$):} By Assumption~\ref{assum:7}, $A(0)$ is primitive. Therefore, from \cite[Theorem~4.1]{kon2005nonexistence}, it follows that every $x(0) \in \partial\mathcal D$ leaves $\partial\mathcal D$ and, after some 
$k^\prime$ ($\in \mathbb{Z}$), enters $\text{int }\mathcal D$. Thereafter, by applying the reasoning in Case 1, it is clear that this orbit $\{x(k)\}_{k \in \mathbb{Z}_{\geq 0}}$ (with $x(0) \in \partial\mathcal D$ and $x(k^\prime) \in \text{int }\mathcal D$) cannot converge to $\partial\mathcal D$. This completes the proof.~\qed

\section{Numerical Examples}\label{sec;sims}
\begin{figure}[h]
\centering
    \includegraphics[width = 7cm, height = 4.5cm]{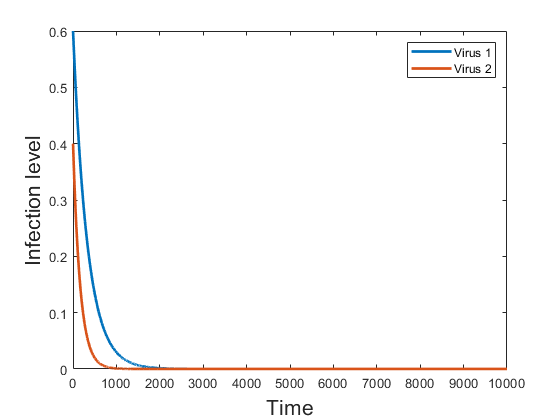}
    \caption{Simulation with two viruses (red and blue), converging to the DFE.}
\label{fig:thm2}
\end{figure}

\seb{We illustrate our results on a fully connected network of ($n=)10$ nodes. Each entry in the matrix $B^1$ (which is the weighted adjacency matrix for the spread of virus~$1$, scaled by the infection rate of each node with respect to virus~$1$) is a random scalar drawn from the uniform distribution in the interval $(0,1)$. We set $B^2=B^1+I_{10 \times 10}$. We choose $D^1=30 \times I$, and $D^2=60 \times I$. We set $h=0.001$. With the aforementioned choice of parameters, it turns out that $\rho(I-hD^1+hB^1)=0.975$, and  $\rho(I-hD^2+hB^2)=0.946$. Therefore, in line with the result in Theorem~\ref{thm:DFE:expo}, virus~$1$ (resp. virus~$2$) gets eradicated exponentially quickly; see blue (resp. red) line in Figure~\ref{fig:thm2}.}

\begin{figure}[h]
\centering
    \includegraphics[width = 7cm, height = 4cm]{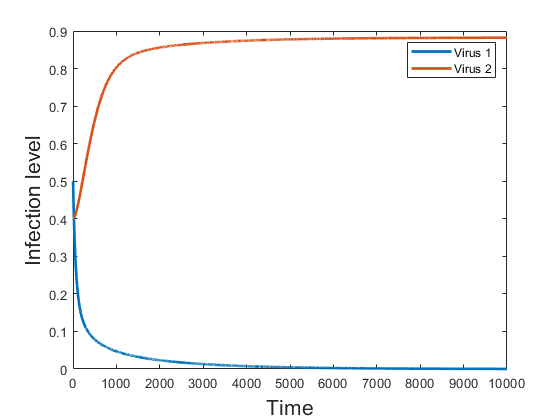}
    \caption{Virus~$1$ dies out, while virus~$2$ becomes endemic.}
    \label{fig:thm3}
\end{figure}

\seb{For the next simulation, we use the same network and the sampling rate as for the simulation in Figure~\ref{fig:thm2}, with the exception that every entry in both $B^1$ and $B^2$ is a random scalar drawn from the uniform distribution in the interval $(0,1)$. Entries in $D^1$ and $D^2$ are also chosen in a similar fashion, except that each element in $D^1$ is multiplied by $20$. We choose $x^1(0)=0.5\times \textbf{1}$, and $x^2(0)=0.4\times \textbf{1}$. With such a choice of parameters, we have that $\rho(I-hD^1+hB^1)=0.9989$, and  $\rho(I-hD^2+hB^2)=1.0045$. Consequently, consistent with the result in Theorem~\ref{thm:boundary:global}, the dynamics of the system converge to the single-virus endemic equilibrium of virus~$2$ (i.e, $\bar{x}^2=[ \begin{smallmatrix}0.842 && 0.83 && 0.98 && 0.85 && 0.88 && 0.91 && 0.89 && 0.82 && 0.96 && 0.89 \end{smallmatrix}]$); see the red line in Figure~\ref{fig:thm3}.}

\begin{figure}[h]
\centering
    \includegraphics[width = 7cm, height = 4cm]{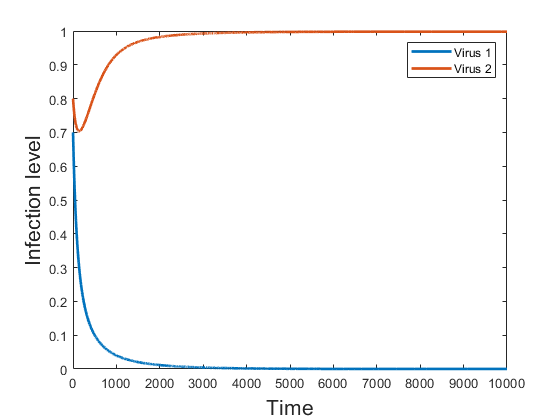}
    \caption{Simulation with two viruses (red and blue).   The single-virus endemic equilibrium of virus~1 is unstable, whereas that of virus~2 is asymptotically stable.}
    \label{fig:thm4}
\end{figure}

\seb{For the next simulation, the setup remains the same as that in the simulation for Figure~\ref{fig:thm3}, with the exception that for a randomly generated choice of $B^1$ and $B^2$, the healing rates are chosen as follows: $D^1 =\diag([\begin{smallmatrix}
    9.19 & 0.9& 2.55& 4.27& 5.77& 8.995 &2.18 &9.67& 4.33& 7.84
\end{smallmatrix}])$, and $D^2=\diag([\begin{smallmatrix}
    0.01 & 0.013 & 0.015& 0.016& 0.06& 0.015& 0.011& 0.0015& 0.005& 0.003
\end{smallmatrix}])$. We choose $x^1(0)=0.7\times \textbf{1}$, and $x^2(0)=0.8\times \textbf{1}$. It turns out that $\rho(I-hD^1+hB^1)=1.0014$, and  $\rho(I-hD^2+hB^2)=1.005$; hence, $\rho(I-hD^1+hB^1)>1$, and  $\rho(I-hD^2+hB^2)>1$. Furthermore, $\rho(I-hD^1+(I-\bar{X}^2)hB^1) = 0.9991$, and
$\rho(I-hD^2+(I-\bar{X}^1)hB^2) = 1.005$; hence $\rho(I-hD^1+(I-\bar{X}^2)hB^1) <1$, and $\rho(I-hD^2+(I-\bar{X}^1)hB^2)>1$. Consequently, in line with our findings in Theorem~\ref{thm:boundary:local}}, the  single-virus endemic equilibrium corresponding to virus~1 is unstable (see blue line in Figure~\ref{fig:thm4}), while the single-virus endemic equilibrium corresponding to virus~2 is asymptotically stable (see the red line in Figure~\ref{fig:thm4}).


\section{Conclusion}\label{sec:conclusion}
The paper dealt with the analysis of the discrete-time networked competitive bivirus SIS model. Specifically, we showed that the system is strongly monotone, and that, under certain assumptions, it does not admit any periodic orbit. We identified a sufficient condition for exponential convergence to the DFE. Thereafter, assuming that only one of the viruses is alive, we identified a sufficient condition for global asymptotic convergence to the endemic equilibrium of this virus - the proof does not depend on the construction of Lyapunov functions. Assuming that both the viruses are alive, we secured a sufficient condition for local asymptotic convergence to the boundary equilibrium of one of the viruses. Finally, we provided a sufficient (resp. necessary) condition for the existence  of a coexistence equilibrium. 

Problems of further interest include, but are not limited to, establishing  counting results for the number of coexistence equilibria; establishing if and when  global convergence to the single-virus endemic equilibrium occurs even when the reproduction number of each virus is larger than one; and devising feedback control strategies for virus mitigation.

\section*{Acknowledgements}
The first author is grateful to Professor Diego Deplano (Department of Electrical and Electronic Engineering, University of Cagliari, Sardegna, Italy) for several in-depth discussions on the result of Theorem~\ref{thm:generic:convergence}.
\bibliography{ReferencesRice}

\begin{thebibliography}{10}
\providecommand{\url}[1]{#1}
\csname url@samestyle\endcsname
\providecommand{\newblock}{\relax}
\providecommand{\bibinfo}[2]{#2}
\providecommand{\BIBentrySTDinterwordspacing}{\spaceskip=0pt\relax}
\providecommand{\BIBentryALTinterwordstretchfactor}{4}
\providecommand{\BIBentryALTinterwordspacing}{\spaceskip=\fontdimen2\font plus
\BIBentryALTinterwordstretchfactor\fontdimen3\font minus
  \fontdimen4\font\relax}
\providecommand{\BIBforeignlanguage}[2]{{%
\expandafter\ifx\csname l@#1\endcsname\relax
\typeout{** WARNING: IEEEtran.bst: No hyphenation pattern has been}%
\typeout{** loaded for the language `#1'. Using the pattern for}%
\typeout{** the default language instead.}%
\else
\language=\csname l@#1\endcsname
\fi
#2}}
\providecommand{\BIBdecl}{\relax}
\BIBdecl

\bibitem{hethcote1991modeling}
H.~W. Hethcote and J.~W. Van~Ark, ``Modeling {HIV} transmission and {AIDS} in
  the {United States},'' \emph{Lecture notes in Biomathematics}, vol.~95, 1991.

\bibitem{hethcote2000mathematics}
H.~W. Hethcote, ``The mathematics of infectious diseases,'' \emph{SIAM Review},
  vol.~42, no.~4, pp. 599--653, 2000.

\bibitem{van2009virus}
P.~Van~Mieghem, J.~Omic, and R.~Kooij, ``Virus spread in networks,''
  \emph{IEEE/ACM Trans. on Networking (TON)}, vol.~17, no.~1, pp. 1--14, 2009.

\bibitem{easley2010networks}
D.~Easley, J.~Kleinberg \emph{et~al.}, \emph{Networks, crowds, and
  markets}.\hskip 1em plus 0.5em minus 0.4em\relax Cambridge University Press,
  2010, vol.~8.

\bibitem{pare2020modeling}
P.~E. Par{\'e}, C.~L. Beck, and T.~Ba{\c{s}}ar, ``Modeling, estimation, and
  analysis of epidemics over networks: An overview,'' \emph{Annual Reviews in
  Control}, vol.~50, pp. 345--360, 2020.

\bibitem{lajmanovich1976deterministic}
A.~Lajmanovich and J.~A. Yorke, ``A deterministic model for gonorrhea in a
  nonhomogeneous population,'' \emph{Mathematical Biosciences}, vol.~28, no.
  3-4, pp. 221--236, 1976.

\bibitem{gracy2020analysis}
S.~Gracy, P.~E. Par{\'e}, H.~Sandberg, and K.~H. Johansson, ``Analysis and
  distributed control of periodic epidemic processes,'' \emph{IEEE Transactions
  on Control of Network Systems}, vol.~8, no.~1, pp. 123--134, 2020.

\bibitem{carlos2}
C.~Castillo-Chavez, W.~Huang, and J.~Li, ``Competitive exclusion and
  coexistence of multiple strains in an {SIS STD} model,'' \emph{SIAM Journal
  on Applied Mathematics}, vol.~59, no.~5, pp. 1790--1811, 1999.

\bibitem{castillo1989epidemiological}
C.~Castillo-Chavez, H.~W. Hethcote, V.~Andreasen, S.~A. Levin, and W.~M. Liu,
  ``Epidemiological models with age structure, proportionate mixing, and
  cross-immunity,'' \emph{Journal of {M}athematical {B}iology}, vol.~27, no.~3,
  pp. 233--258, 1989.

\bibitem{sahneh2014competitive}
F.~D. Sahneh and C.~Scoglio, ``Competitive epidemic spreading over arbitrary
  multilayer networks,'' \emph{Physical Review E}, vol.~89, no.~6, p. 062817,
  2014.

\bibitem{liu2019analysis}
J.~Liu, P.~E. Par{\'e}, A.~Nedi{\'c}, C.~Y. Tang, C.~L. Beck, and
  T.~Ba{\c{s}}ar, ``Analysis and control of a continuous-time bi-virus model,''
  \emph{IEEE Transactions on Automatic Control}, vol.~64, no.~12, pp.
  4891--4906, 2019.

\bibitem{ye2021convergence}
M.~Ye, B.~D.~O. Anderson, and J.~Liu, ``Convergence and equilibria analysis of
  a networked bivirus epidemic model,'' \emph{SIAM Journal on Control and
  Optimization}, vol.~60, no.~2, pp. S323--S346, 2022.

\bibitem{pare2020analysis}
P.~E. Par{\'e}, D.~Vrabac, H.~Sandberg, and K.~H. Johansson, ``Analysis, online
  estimation, and validation of a competing virus model,'' in \emph{2020
  American Control Conference (ACC)}.\hskip 1em plus 0.5em minus 0.4em\relax
  IEEE, 2020, pp. 2556--2561.

\bibitem{pare2021multi}
P.~E. Par{\'e}, J.~Liu, C.~L. Beck, A.~Nedi{\'c}, and T.~Ba{\c{s}}ar,
  ``Multi-competitive viruses over time-varying networks with mutations and
  human awareness,'' \emph{Automatica}, vol. 123, p. 109330, 2021.

\bibitem{liu2020stability}
F.~Liu, C.~Shaoxuan, X.~Li, and M.~Buss, ``On the stability of the endemic
  equilibrium of a discrete-time networked epidemic model,''
  \emph{IFAC-PapersOnLine}, vol.~53, no.~2, pp. 2576--2581, 2020.

\bibitem{anderson2023equilibria}
B.~D. Anderson and M.~Ye, ``Equilibria analysis of a networked bivirus epidemic
  model using {P}oincar\'e-{H}opf and manifold theory,'' \emph{SIAM Journal on
  Applied Dynamical Systems}, {to appear}.

\bibitem{atkinson1991introduction}
K.~Atkinson, \emph{An introduction to numerical analysis}.\hskip 1em plus 0.5em
  minus 0.4em\relax John wiley \& sons, 1991.

\bibitem{hirsch2005monotone}
M.~W. Hirsch and H.~Smith, ``Monotone maps: a review,'' \emph{Journal of
  Difference Equations and Applications}, vol.~11, no. 4-5, pp. 379--398, 2005.

\bibitem{deplano2020nonlinear}
D.~Deplano, M.~Franceschelli, and A.~Giua, ``A nonlinear perron--frobenius
  approach for stability and consensus of discrete-time multi-agent systems,''
  \emph{Automatica}, vol. 118, p. 109025, 2020.

\bibitem{hirsch2006attractors}
M.~W. Hirsch, ``Attractors for discrete-time monotone dynamical systems in
  strongly ordered spaces,'' in \emph{Geometry and Topology: Proceedings of the
  Special Year held at the University of Maryland, College Park
  1983--1984}.\hskip 1em plus 0.5em minus 0.4em\relax Springer, 2006, pp.
  141--153.

\bibitem{sontag2007monotone}
E.~D. Sontag, ``Monotone and near-monotone biochemical networks,''
  \emph{Systems and synthetic biology}, vol.~1, no.~2, pp. 59--87, 2007.

\bibitem{hirsch2020positive}
M.~W. Hirsch, ``Positive equilibria and convergence in subhomogeneous monotone
  dynamics,'' in \emph{Comparison methods and stability theory}.\hskip 1em plus
  0.5em minus 0.4em\relax CRC Press, 2020, pp. 169--188.

\bibitem{cui2022discrete}
S.~Cui, F.~Liu, H.~Jard{\'o}n-Kojakhmetov, and M.~Cao, ``Discrete-time
  layered-network epidemics model with time-varying transition rates and
  multiple resources,'' \emph{arXiv preprint arXiv:2206.07425}, 2022.

\bibitem{sebin:ifac23}
\BIBentryALTinterwordspacing
S.~Gracy, Y.~Wang, P.~E. Par\'e, and C.~Uribe, ``Multi-competitive virus spread
  over a time-varying networked {SIS} model with an infrastructure network,''
  in \emph{IFAC World Congress}.\hskip 1em plus 0.5em minus 0.4em\relax IFAC,
  2023, {N}ote: To Appear. [Online]. Available:
  \url{https://arxiv.org/pdf/2303.08859.pdf}
\BIBentrySTDinterwordspacing

\bibitem{rantzer2011distributed}
A.~Rantzer, ``Distributed control of positive systems,'' in \emph{Proceedings
  of the 50th IEEE Conference on Decision and Control and European Control
  Conference}, 2011, pp. 6608--6611.

\bibitem{horn2012matrix}
R.~A. Horn and C.~R. Johnson, \emph{Matrix Analysis}.\hskip 1em plus 0.5em
  minus 0.4em\relax Cambridge University Press, 2012.

\bibitem{axel2020TAC}
\BIBentryALTinterwordspacing
A.~Janson, S.~Gracy, P.~E. Par\'e, H.~Sandberg, and K.~H. Johansson,
  ``Networked multi-virus spread with a shared resource: Analysis and
  mitigation strategies,'' \emph{https://arxiv.org/pdf/2011.07569.pdf}, 2020.
  [Online]. Available: \url{https://arxiv.org/pdf/2011.07569.pdf}
\BIBentrySTDinterwordspacing

\bibitem{vidyasagar2002nonlinear}
M.~Vidyasagar, \emph{Nonlinear Systems Analysis}.\hskip 1em plus 0.5em minus
  0.4em\relax Siam, 2002, vol.~42.

\bibitem{dancer1998some}
E.~Dancer, ``Some remarks on a boundedness assumption for monotone dynamical
  systems,'' \emph{Proceedings of the American Mathematical Society}, vol. 126,
  no.~3, pp. 801--807, 1998.

\bibitem{varga1999matrix}
\BIBentryALTinterwordspacing
R.~Varga, \emph{Matrix Iterative Analysis}, ser. Springer Series in
  Computational Mathematics.\hskip 1em plus 0.5em minus 0.4em\relax Springer
  Berlin Heidelberg, 1999. [Online]. Available:
  \url{https://books.google.se/books?id=U2XYs1DyKiYC}
\BIBentrySTDinterwordspacing

\bibitem{qu2009cooperative}
Z.~Qu, \emph{Cooperative control of dynamical systems: applications to
  autonomous vehicles}.\hskip 1em plus 0.5em minus 0.4em\relax Springer Science
  \& Business Media, 2009.

\bibitem{hess1991stability}
P.~Hess and E.~Dancer, ``Stability of fixed points for order-preserving
  discrete-time dynamical systems.'' 1991.

\bibitem{ben:pre}
\BIBentryALTinterwordspacing
M.~Ye, B.~D.~O. Anderson, A.~Janson, S.~Gracy, and K.~H. Johansson,
  ``Competitive epidemic networks with multiple survival-of-the-fittest
  outcomes,'' \emph{Systems \& Control Letters}, 2023, {N}ote: {U}nder
  {R}eview. [Online]. Available: \url{https://arxiv.org/pdf/2111.06538.pdf}
\BIBentrySTDinterwordspacing

\bibitem{FB-LNS}
\BIBentryALTinterwordspacing
F.~Bullo, \emph{Lectures on Network Systems}, {1.6}~ed.\hskip 1em plus 0.5em
  minus 0.4em\relax Kindle Direct Publishing, 2022. [Online]. Available:
  \url{https://fbullo.github.io/lns}
\BIBentrySTDinterwordspacing

\bibitem{kon2005nonexistence}
R.~Kon, ``Nonexistence of synchronous orbits and class coexistence in matrix
  population models,'' \emph{SIAM Journal on Applied Mathematics}, vol.~66,
  no.~2, pp. 616--626, 2005.

\end{thebibliography}
\end{document}